\documentclass[epsfig,graphics,twocolumn,showpacs,floatfix,mathbbm,prl,superscriptaddress]{revtex4}
\usepackage{graphicx}
\usepackage{color}
\usepackage{ulem}
\newcommand{\PrYSO}{Pr$^{3+}$:Y$_2$Si{O$_5$}}
\newcommand{\YSO}{Y$_2$Si{O$_5$}}

\begin{document}

\title[a]{Quantum Storage of Heralded Single Photons in a Praseodymium Doped Crystal}%

\author{Daniel Riel\" ander}
\affiliation{ICFO-Institut de Ciencies Fotoniques, Mediterranean Technology Park, 08860 Castelldefels (Barcelona), Spain}%

\author{Kutlu Kutluer}
\affiliation{ICFO-Institut de Ciencies Fotoniques, Mediterranean Technology Park, 08860 Castelldefels (Barcelona), Spain}%

\author{Patrick M. Ledingham}
\affiliation{ICFO-Institut de Ciencies Fotoniques, Mediterranean Technology Park, 08860 Castelldefels (Barcelona), Spain}%

\author{Mustafa G\"{u}ndo\u{g}an}
\affiliation{ICFO-Institut de Ciencies Fotoniques, Mediterranean Technology Park, 08860 Castelldefels (Barcelona), Spain}%

\author{Julia Fekete}
\altaffiliation{Present address: Physics Department, University of Otago, New Zealand}
\affiliation{ICFO-Institut de Ciencies Fotoniques, Mediterranean Technology Park, 08860 Castelldefels (Barcelona), Spain}%

\author{Margherita Mazzera}
\email{margherita.mazzera@icfo.es}
\affiliation{ICFO-Institut de Ciencies Fotoniques, Mediterranean Technology Park, 08860 Castelldefels (Barcelona), Spain}%

\author{Hugues de Riedmatten}
\affiliation{ICFO-Institut de Ciencies Fotoniques, Mediterranean Technology Park, 08860 Castelldefels (Barcelona), Spain}%
\affiliation{ICREA-Instituci\'{o} Catalana de Recerca i Estudis Avan\c cats, 08015 Barcelona, Spain}%

\date{\today}

\begin{abstract}
We report on experiments demonstrating the reversible mapping of
heralded single photons to long lived collective optical atomic
excitations stored in a {\PrYSO} crystal. A cavity-enhanced
spontaneous down-conversion source is employed to produce widely
non-degenerate narrow-band ($\approx 2\,\mathrm{MHz}$) photon-pairs. The idler photons, whose frequency is
compatible with telecommunication optical fibers, are used to
herald the creation of the signal photons, compatible with the
Pr$^{3+}$ transition. The signal photons are stored and retrieved
using the atomic frequency comb protocol.
We demonstrate storage times up to $4.5\,\mathrm{\mu s }$ while
preserving non-classical correlations between the heralding and the
retrieved photon. This is more than 20 times longer than in
previous realizations in solid state devices, and implemented in a
system ideally suited for the extension to spin-wave storage.

\end{abstract}

\maketitle

Many protocols in quantum information science rely on the
efficient and reversible interaction between photons and matter
\cite{Hammerer2010}. The interaction lays the basis for the
realization of quantum memories for light and of their
application, e.g. in quantum repeaters
\cite{Briegel1998,Sangouard2011}. Possible choices for the system
used to store light are single atoms in cavities
\cite{Specht2011}, cold or hot atomic gases
\cite{Chaneliere2005,Chou2005,Simon2007c,Radnaev2010,Zhang2011,Julsgaard2004,Eisaman2005,
Reim2011,Hosseini2011}, or rare earth (RE) doped solid state
systems \cite{Tittel2010}. Thanks to the weak interaction between
the optical active ions and the environment, RE doped crystals
offer, when cryogenically cooled, the long optical and spin
coherence times typical of atomic systems, free of the drawbacks
deriving from atomic motion \cite{Macfarlane2002}. Moreover they
possess the benefits of the solid state systems, such as strong
interaction with light, allowing for efficient storage of photons
\cite{Hedges2010,Sabooni2013} and prospect for integrated devices.
Furthermore, their inhomogeneously broadened absorption lines can
be tailored in appropriate structures, like atomic frequency combs
(AFCs), to enable storage protocols with remarkable properties
(e.g. temporal or frequency multiplexing)
\cite{Riedmatten2008,Afzelius2009,Usmani2010,Bonarota2011,Gundogan2013,Afzelius2010,Sinclair2013}.

Single photon level weak coherent pulses
\cite{Sabooni2010,Chaneliere2010} and qubits
\cite{Riedmatten2008,Gundogan2012,Zhou2012} have been stored in
the excited state of rare-earth doped crystals using the AFC
scheme. This has recently been extended to the ground state, in
the regime of a few photons per pulse \cite{Timoney2013}. The
storage of non-classical light generated by spontaneous parametric
down-conversion (SPDC) has also been demonstrated and enabled
entanglement between one photon and one collective optical atomic
excitation in a crystal \cite{Clausen2011,Saglamyurek2011},
entanglement between two crystals \cite{Usmani2012}, and single
photon qubit storage \cite{Clausen2012,Saglamyurek2012}. However,
the mapping of non-classical light using AFC in rare earth doped
crystals was obtained so far only in systems with two ground state
levels, thus inherently limited to the optical coherence and not
directly extendable to spin-wave storage.

On the contrary, Pr$^{3+}$ or Eu$^{3+}$ doped crystals have the
required level structure for spin-wave storage
\cite{Afzelius2010,Gundogan2013,Timoney2013}.

In particular, {\PrYSO} is one of the optical memories with the best
demonstrated properties. Storage efficiencies as high as 69 $\%$
for weak coherent states \cite{Hedges2010} and storage times up to
1 minute (the longest in any system so far) for classical images
\cite{Heinze2013} have been reported. Despite these extraordinary
performances, which make {\PrYSO} an excellent candidate for
quantum memories, the storage of quantum light has never been
achieved in this material. As a matter of fact, the
pseudoquadrupolar interaction which splits the crystal-field
singlets into hyperfine sub-levels, providing the three-fold
ground state required for the storage in the spin state, also
establishes a tight bound ($< 4\, \mathrm{MHz}$) to the bandwidth
of the single photons to be stored. Recently a SPDC source has
been developed to create ultranarrow-band photon-pairs, with the
signal and the idler photons compatible with the {Pr$^{3+}$}
transition at $606\,\mathrm{nm}$ and with telecommunication
optical fibers, respectively \cite{Fekete2013}.

In this Letter we report on experiments where one photon of the
pair (signal), whose presence is heralded by the other photon
(idler), is stored as a collective optical atomic excitation in a
{\PrYSO} crystal with the AFC protocol. We show that the
non-classical correlation between the two photons is preserved for
storage times up to $4.5\,\mathrm{\mu s }$, more than 20 times
longer than in previous solid state experiments
\cite{Clausen2011,Saglamyurek2011}. The demonstrated non-classical
correlations between an atomic excitation stored in a crystal and
a photon at telecommunication wavelengths are an essential
resource to generate heralded entanglement between remote crystals
\cite{Simon2007a}.

The AFC scheme \cite{Afzelius2009} relies on the creation of a
series of narrow  absorbing peaks with periodicity $\Delta$ in a transparency window created within the
inhomogeneous absorption profile of the crystal. The single photon
is then mapped onto the crystal, leading, in the ideal case, to a
single collective optical excitation:
$\sum_{j=1}^{N_A}e^{-i\overrightarrow{k_p} \cdot \overrightarrow{x_j}}e^{-i\delta_jt}|g_1\cdot\cdot\cdot
e_j\cdot\cdot\cdot g_{N_A}\rangle$, where $N_A$ is the number of atoms, $|g\rangle$ and
$|e\rangle$ are the ground and excited state, respectively,
$\overrightarrow{k_p}$ is the single photon wave vector, and
$\overrightarrow{x_j}$ ($\delta_j$) is the position (detuning) of
atom $j$. After an initial inhomogeneous dephasing, the atoms will
rephase after a time $\tau = 1/{\Delta}$ giving rise to a
re-emission of the photon in the forward direction, the so called
AFC echo \cite{Riedmatten2008}.

\begin{figure}
   \centering
   \includegraphics[width=.48\textwidth]{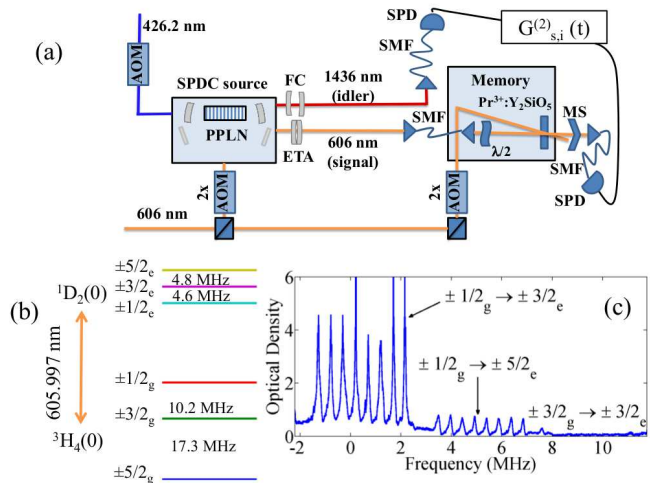}
   \caption{(Color online) (a) Experimental setup. AOM, acousto-optic modulator
(2x when double pass); FC, filter cavity; ETA, etalon; SMF,
single-mode fiber; MS, mechanical shutter; SPD, single photon
detector; $\lambda/2$, half-wave plate; (b) Energy level scheme of
{Pr$^{3+}$} in {\YSO} restricted to the fundamental levels of the
ground $^{3}$H$_{4}$ and the excited $^{1}$D$_{2}$ manifolds. (c)
Example of AFC with $\Delta = 500\,\mathrm{kHz}$.}
   \label{setup}
   \end{figure}

Figure \ref{setup}(a) represents the experimental setup. The
coherent light at $606\,\mathrm{nm}$ is obtained by sum frequency
generation of $1570\,\mathrm{nm}$ and $987\,\mathrm{nm}$ lasers
\cite{Gundogan2013}. Its frequency is locked to a temperature
stabilized cavity placed in a home-made vacuum chamber. Two beams
are spatially separated, the first being used as a frequency
reference for the photon-pair source \cite{Fekete2013} and the
second for the memory preparation \cite{Gundogan2013}. Their
frequency and amplitude are varied by double-pass acousto-optic
modulators (AOMs) driven by an arbitrary waveform generator
(Signadyne). After the AOMs, the beams are coupled into
polarization maintaining single-mode fibers and out-coupled in two
separated optical benches hosting the source and the cryostat
(closed cycle cryogenic cooler, Oxford V14), which is exploited to
cool the {\PrYSO} crystal down to $2.8\,\mathrm{K}$.

The heralded single photons are obtained from the photon-pair
source, whose detailed description is given in ref.
\cite{Fekete2013}. A CW pump beam at $426.2\,\mathrm{nm}$,
modulated by an AOM, produces photons at $606\,\mathrm{nm}$
(signal) and $1436\,\mathrm{nm}$ (idler) by SPDC in a
temperature-stabilized periodically poled lithium niobate (PPLN)
crystal. A bow-tie cavity surrounding the crystal enhances the
SPDC process for the resonant frequencies. The cavity is
stabilized with the reference beam at $606\,\mathrm{nm}$, to
ensure the signal resonance with the {\PrYSO} crystal. To maintain
the cavity resonant with the idler, the pump frequency is locked
using light at the idler frequency created by difference frequency
generation of the pump and the $606\,\mathrm{nm}$ reference beam
passing through the same PPLN crystal. The double resonance allows
a very efficient suppression of the redundant spectral modes due
to the so called clustering effect \cite{Fekete2013}. The
photon-pairs' spectrum is thus composed of a main- and two
secondary-clusters separated by $44.5\,\mathrm{GHz}$, each
containing four longitudinal modes \cite{Fekete2013}. In order to
operate in the single-mode regime, the heralding telecom photons
are filtered by a home-made filter cavity (linewidth $\approx
80\,\mathrm{MHz}$ and free spectral range (FSR) $\approx 16.8\,\mathrm{GHz}$). They are
then coupled into a single mode fiber and detected  by an InGaAs
single photon detector (SPD, IdQuantique id220, detection
efficiency $\eta_{d,i}=10\,\%$ and $400\,\mathrm{Hz}$ dark count
rate). The heralded $606\,\mathrm{nm}$ photons are filtered with
an etalon (linewidth $\approx 10\,\mathrm{GHz}$ and FSR $\approx
60\,\mathrm{GHz}$), resulting in the mere suppression of the
secondary-clusters. They are then sent to the crystal via a single
mode fiber. The optical transmission of the signal (idler) photon
from the output of the bow-tie cavity to the cryostat (SPD) is
$\eta_s=0.18$ ($\eta_i=0.22$). To optimize the detection of the
stored and retrieved photons, we switch off the pump beam using
the AOM after the detection of an idler photon, thus interrupting
the creation of new photon-pairs. The correlation time of the
photon-pair is measured from the idler-signalcross-correlation
function $G^{(2)}_{s,i} (t)$ to be $\tau_{c} = 108\,\mathrm{ns}$,
corresponding to a photon bandwidth of $\delta\nu = 2.3
\,\mathrm{MHz}$ ($1.84 \,\mathrm{MHz}$) for the signal (idler)
photons \cite {Fekete2013}.

Our storage device is a $3\,\mathrm{mm}$ thick {\YSO} sample doped
with a Pr$^{3+}$ concentration of $0.05\,\%$. The relevant optical
transition is at $605.977\,$nm with a measured absorption
coefficient of $\alpha = 23\,\mathrm{cm}^{-1}$ and an
inhomogeneous linewidth of $5\,\mathrm{GHz}$ \cite{Gundogan2013}.
A half-wave plate ($\lambda/2$) ensures the polarization of the
photons to be aligned close to the optical D$_2$ axis of the
crystal, in order to maximize the absorption. To prevent noise
from the strong preparation beam polluting the single photon mode,
we use two different optical paths with an angle of
$2.5\,\mathrm{degrees}$. The maximum power in the preparation mode
before the cryostat window is $6\,\mathrm{mW}$ and the beam
diameter at the crystal is $150\,\mu\mathrm{m}$. The beam diameter
for the input mode is $50\,\mu\mathrm{m}$. The single photon mode
is directed to a Si-SPD (Excelitas Technologies, efficiency
$\eta_{d,s} = 32\,\%$ and $10\,\mathrm{Hz}$ dark counts) via a
single mode fiber.

To prepare the memory we follow the procedure thoroughly described
in \cite{Nilsson2004,Gundogan2013}. We prepare a transparency
window within the inhomogeneously broadened Pr$^{3+}$ absorption
at $606\,$nm by sweeping the laser frequency by
$12\,\mathrm{MHz}$. The narrow transparency window contributes to
select only one mode among the four remaining in the main-cluster
of the source spectrum which are separated by $412\,\mathrm{MHz}$
\cite{Fekete2013}. Afterwards we tailor a single class AFC on the
$\pm {1}/{2}_\textrm{g}-\pm {3}/{2}_\textrm{e}$ transition by
first burning back atoms with pulses at frequencies differing by
$\Delta$ and then performing a cleaning sweep in the region of the
$\pm{3}/{2}_\textrm{g}-\pm{3}/{2}_\textrm{e}$ transition (see Fig.
\ref{setup}(b)). To ensure the efficient absorption of the photons
by the periodic structure, the total comb-width is set to
$3.5\,\mathrm{MHz}$ (see Fig. \ref{setup}(c) for an example with
$\Delta = 500\,\mathrm{kHz}$). With the present preparation
procedure (which lasts for $300\,\mathrm{ms}$) we are able to tailor spectral features as narrow as
$60\,\mathrm{kHz}$. The input photon frequency is selected to be
resonant with the $ \pm {1}/{2}_\textrm{g}- \pm
{3}/{2}_\textrm{e}$ transition. Note that the procedure used for
the comb preparation empties the $\pm{3}/{2}_\textrm{g}$ ground
state. The $\pm{3}/{2}_\textrm{g}-\pm{3}/{2}_\textrm{e}$
transition could then be directly used to transfer optical
excitations to spin excitations, as demonstrated in
\cite{Afzelius2010,Gundogan2013} for bright pulses.

The arrival times of the photons to the detectors are recorded
with a time-stamping card (Signadyne) and used to reconstruct the
second-order cross-correlation function between signal and idler
$G^{(2)}_{s,i} (t)$. A figure of merit for the non-classical
nature of the photon correlations is the normalized
cross-correlation function \cite{Sangouard2011,Clausen2011}
\begin{equation}
g_{s,i}^{(2)} = \frac{p_{s,i}}{p_{s}p_{i}}, \\
\end{equation}

where $p_{s}$ ($p_{i}$) is the probability to detect a signal
(idler) photon and $p_{s,i}$ the probability to detect a
coincidence in a time window $\Delta t_d = 400\, \mathrm{ns}$
centered at zero time delay. All the results presented in this
paper are from raw data, without any background or dark count
subtraction, unless otherwise stated.

\begin{figure}
   \centering
   \includegraphics[width=.48\textwidth]{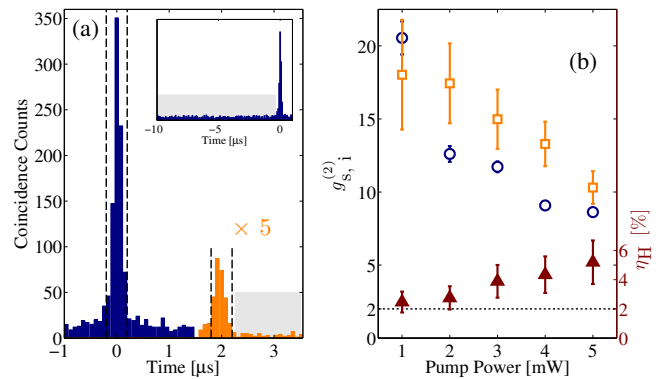}
   \caption{(Color online) (a) $G^{(2)}_{s,i} (t)$ histogram without (blue) and
with (orange) AFC. The preprogrammed storage time is
$2\,\mathrm{\mu s }$ and the power of the $426.2\,\mathrm{nm}$
pump is $2\,\mathrm{mW}$. The time-bin size is $5\,\mathrm{ns}$.
The dashed lines and shaded area define the detection windows for
the coincidences ($\Delta t_d = 400\, \mathrm{ns}$) and noise,
respectively, used to compute $g^{(2)}_{s,i}$. (b) The
$g^{(2)}_{s,i}$ values as a function of the pump power for the AFC
echo (empty squares) are compared to those for the input photons
(empty circles). The heralding efficiencies $\eta_{H}$ are also
reported (filled triangles).The integration times range between
$30\,$ and $60\,\mathrm{minutes}$. The error bars are evaluated
from the raw number of counts assuming Poissonian statistics. The
dotted line corresponds to the classical limit $g^{(2)}_{s,i} = 2$
for two-mode squeezed states. }
   \label{g2vsPP}
   \end{figure}

We first characterize the input heralded single photon by sending
it through the memory crystal when only a transparency window is
created (see blue histogram centered at zero time delay in Fig.
\ref{g2vsPP}(a)). The values of $g^{(2)}_{s,i}$ as a function of
the pump power  are plotted in Fig. \ref{g2vsPP}(b). Despite the
expected decrease for increased pump powers due to the production
of multiple pairs \cite{Foertsch2012}, the values remain well above the
classical limit of 2 (dotted line) for two-mode squeezed states.

In order to unambiguously confirm the non-classical nature of the
correlation, we also measure the marginal auto-correlation
function for the signal and idler mode, $g^{(2)}_{s,s}$ and
$g^{(2)}_{i,i}$. We find, at pump
power of $5\,\mathrm{mW}$, $g^{(2)}_{s,s} = 1.14 \pm 0.03$  and
$g^{(2)}_{i,i} = 1.07 \pm 0.02$, leading to strong violation of
the Cauchy-Schwarz inequality in the form $ R =
{(g^{(2)}_{s,i})^2} / ({{g^{(2)}_{i,i} \times g^{(2)}_{s,s}}) } =
61 \pm 2 \not\leq 1$ which proves non-classical correlations
without any assumption on the created state (see Supplemental Material).

The detected coincidence rate (within $\Delta t_d$) is $C_d =
(0.83\pm 0.14)\,\mathrm{Hz/mW}$. From this value, we infer the
generated coincidence rate $C_g$ at the output of the source
cavity
$C_g=C_d/(\eta_i\cdot\eta_{d,i}\cdot\eta_s\cdot\eta_{loss}\cdot\eta_{d,s})=2.8\,\mathrm{kHz/mW}$,
where $\eta_{loss}=0.225$ is the transmission from the input of
the cryostat to the signal SPD, including the duty cycle of the
memory (50 $\%$, see Supplemental Material). We can also evaluate the heralding efficiency in
front of the cryostat, $\eta_{H} = {p_{s,i}}/ (p_{i} \times
\eta_{d,s} \times \eta_{loss}) $, whose values as a function of
pump power are plotted in Fig. \ref{g2vsPP}(b) (full triangles).
Note that $\eta_H$ is mainly limited by $\eta_{s}$ and dark counts
in the heralding SPD. Subtracting the contribution of the
$1436\,\mathrm{nm}$ detector dark counts we find corrected
heralding efficiencies almost constant, ${\eta_{H}}^{DC} \approx
6.3\,\%$ ($35\,\%$ at the output of the source cavity), over the
whole range of pump powers investigated. Finally, we estimate that
more than $95\,\%$ of the heralded signal photons detected after
the crystal are resonant with the atoms (see Supplemental
Material) \cite{Wolfgramm2011}.

Once the non-classical nature of the input photons propagating
through the transparency window is demonstrated, we prepare the
AFC and reconstruct the $G^{(2)}_{s,i} (t)$ function for the
stored and retrieved photons. Figure \ref{g2vsPP}(a) includes the
coincidence histogram when an AFC is producing a collective
re-emission at a delay of $\tau = 2\, \mu s$ (orange trace). The
values of $g^{(2)}_{s,i}$ for the AFC echo as a function of the
pump power are reported in Fig. \ref{g2vsPP}(b) (empty squares).
The count rate in the region of the AFC echo is $C =
(0.043\pm0.03)\,\mathrm{Hz/mW}$. We observe that the echoes
exhibit $g^{(2)}_{s,i}$ values higher than the input photons. We
attribute this unexpected effect to the fact that the AFC acts as
a temporal filter \cite{McAuslan2012} for non-resonant noise
arising from the SPDC source. Since the SPDC pump laser is turned
off after the detection of an idler photon, the AFC delays the
signal to a region free of broadband noise, thus increasing the
$g^{(2)}_{s,i}$. A more quantitative analysis of this effect is
presented in the Supplemental Material. When the pump power
decreases, the detection of the echo is limited by the detector
dark counts. This gives rise to a saturation of the echo
$g^{(2)}_{s,i}$ values which hides the filtering effect of the
storage.

   \begin{figure}
   \centering
  \includegraphics[width=.48\textwidth]{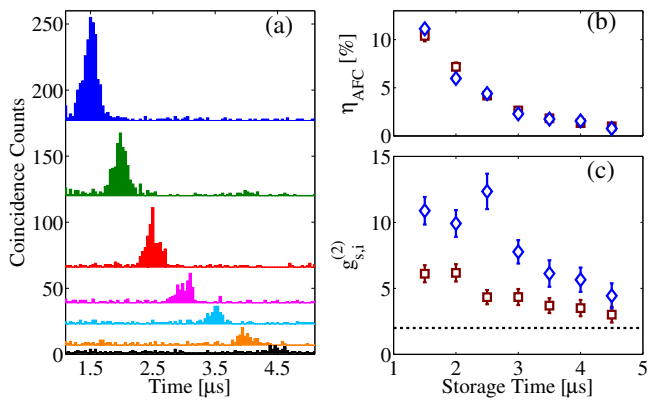}
   \caption{(Color online) (a) AFC echoes at different predetermined storage times observed in coincidence histograms measured at $2\,\mathrm{m W}
   $ pump power (integration time is $40\, \mathrm{minutes}$ and  the time-bin size is
   $5\,\mathrm{ns}$).
The curves are vertically shifted for clarity. (b) and (c) AFC
efficiency and $g^{(2)}_{s,i}$ (calculated in a $400\,
\mathrm{ns}$ window) of the AFC echo as a function of the storage
time. Diamonds: filter cavity in the heralding photon path;
squares: filter cavity removed from the heralding photon path. The
error bars are evaluated from the raw counts assuming Poissonian
statistics. The dotted line corresponds to the classical limit
$g^{(2)}_{s,i} = 2$ for two-mode squeezed states.}
   \label{AFCvsTau}
   \end{figure}

In view of applications in temporally multi-mode quantum memories
with on-demand read-out, the pre-programmed delay $\tau$ must be
tunable to allow the application of control pulses transferring
the excitation to the spin states \cite{Gundogan2013}. Thus, we
changed the spectral periodicity $\Delta$ of the AFC to obtain
increased delays. Figure \ref{AFCvsTau}(a) shows the coincidence
histograms in the region of the AFC echo occurring at different
predetermined storage times. The AFC storage efficiency,
$\eta_{AFC}$, is estimated by comparing the number of counts of
the input propagating through the transparency window and the
echo; it decreases for longer delays, as shown in Fig.
\ref{AFCvsTau}(b), denoting a reduction of the comb finesse
\cite{Riedmatten2008}. We note that the AFC efficiency with single
photons does not decrease with respect to the storage of bright or
weak coherent pulses of similar durations (the different cases
together with a theoretical evaluation are compared in the
Supplemental Material). This confirms that there is no significant
frequency jitter for the single photons. We could reach AFC
storage times of $4.5\,\mathrm{\mu s}$, while preserving
non-classical correlations (see Fig. \ref{AFCvsTau}(c)). As we
stop the production of pairs after the detection of a telecom
photon, the region where the echo lies should not be affected by
unconditional noise coming from the source (see Supplemental
Material). Thus, despite the reduction in the AFC efficiency, the
echo should exhibit constant values of $g^{(2)}_{s,i}$ for
increasing storage times. As a matter of fact, the decrease in the
$g^{(2)}_{s,i}$ values while increasing the storage time is less
pronounced than the drop of the AFC efficiency (compare the
diamonds in panels (b) and (c) of Fig. \ref{AFCvsTau}) and we
attribute the reduction to the limitation given by the detector
dark counts.

We also report the values of $\eta_{AFC}$ and $g^{(2)}_{s,i}$
measured without the filter cavity in the idler arm. The
efficiencies in the two cases agree within the experimental error.
On the contrary, despite the factor $2\,$ gain in the raw number
of counts thanks to the reduced passive losses, the
$g^{(2)}_{s,i}$ without filter cavity decreases due to the
frequency multimodality of the source \cite{Fekete2013}. In fact,
among the four modes coexisting in the main-cluster, only one is
resonant with the AFC. The non-resonant modes in the idler arm
join the start signals for the coincidence histogram without
corresponding stops (the Pr$^{3+}$ absorption acts as a filter for
the non-resonant mode in the $606\,\mathrm{nm}$ arm), thus
decreasing the heralding efficiency and contributing to the noise
increase. Note that frequency multiplexed quantum light storage
could be obtained by creating several AFCs at different
frequencies separated by the SPDC cavity FSR \cite{Sinclair2013}.

Our experiment could be extended to the storage of entangled
qubits (e.g. entangled in energy-time \cite{Clausen2011} or in
polarization with a suitable modification of the pair source).
Alternatively, by doubling the setup, the demonstrated light
matter quantum correlations would enable heralded entanglement
between remote crystals \cite{Simon2007a}. For the two cases, the
visibility $V$ of the two-photon interference would be given by the
measured $g^{(2)}_{s,i}$ of the stored and retrieved photons as
(assuming uncorrelated background):
$V=(g^{(2)}_{s,i}-1)/(g^{(2)}_{s,i}+1$)
\cite{Riedmatten2006,Laurat2007} . For the values of
$g^{(2)}_{s,i}$ measured in this work, this would lead to
visibilities between $0.67$ and $0.9$. In addition, for the latter
experiment, the demonstrated storage time would allow entanglement
between crystals separated by km range distances.

In conclusion, we have demonstrated the reversible mapping of
heralded single photons to collective optical atomic excitations
in a praseodymium doped crystal. We observed an increase of
non-classical correlations between signal and idler photons during
the storage, thanks to a temporal filtering effect due to the AFC.
Storage times up to $4.5\,\mathrm{\mu s }$ while preserving
quantum correlations were observed, more than 20 times longer than
previous solid state experimental realizations.  Furthermore, the
transition used would readily allow the transfer of the excitation
to the ground state to obtain long-lived spin-wave storage of
quantum state of light, provided that the noise induced by the
control beams can be sufficiently reduced.

We thank M. Afzelius and F. Bussi\`eres for interesting discussions.
We acknowledge financial support by the European projects
CHIST-ERA QScale and FP7-CIPRIS (MC ITN-287252), by the ERC
Starting grant QuLIMA and by the Spanish MINECO OQISAM project
(FIS2012-37569). MM acknowledges the Beatriu de Pin\'os program
(2010BP B0014) for financial support.

\hrulefill
\newpage

\hrulefill

\appendix

\section{Supplemental Material}

\maketitle

We report, in the present supplemental material, details about the experimental setup (Section I.),
the characterization of the AFC storage with bright pulses and weak coherent states (Section II.), the second-order auto-correlation measurements of signal and idler photons (Section III.), and the second-order cross-correlation measurements between signal and idler as a function of the signal photons polarization (Section IV.).

\section{Setup}

In the present section we provide additional details about the setup and the measurements. 

The coherent light at $606\,\mathrm{nm}$ is obtained using a periodically poled KTiOPO$_4$ waveguide (AdVR, Inc.) by means of sum frequency generation of $1570\,\mathrm{nm}$ and $987\,\mathrm{nm}$ light. The sources for these are amplified diode lasers (Toptica, DL 100 pro and TA PRO, respectively) \cite{Gundogan2013}. The pump beam for the SPDC source comes from a laser at $426.2\,\mathrm{nm}$ (Toptica, TA SHG).

\begin{table}[htdp]
\begin{center}
\caption{Free spectral range (FSR) and finesse ($\mathcal{F}$) of the cavities present in the esperimental setup.}
 \label{tab:cavities}

\begin{tabular}{|c|c|c|}
\hline  Cavity & FSR (GHz) & $\mathcal{F}$  \\

\hline

$606\,\mathrm{nm}$ laser lock & $1 \,$ & $\approx787$  \\
SPDC & $0.412$ & $\approx 200$ \\
Filter cavity, $1436\,\mathrm{nm}$ & $16.8$ & $\approx 210$ \\
Etalon, $606\,\mathrm{nm}$ & $60$ & $\approx 6$ \\

 \hline
 \end{tabular}
\end{center}
\end{table}

Table \ref{tab:cavities} summarizes the characteristics of the cavities included in the setup. A Fabry-Perot cavity is used to lock the frequency of the $606\,\mathrm{nm}$ laser with the Pound-Drevel-Hall technique. It is assembled in a home-made temperature stabilized vacuum chamber ($2 \times 10^{-7}\,\mathrm{mbar}$). A bow-tie cavity surrounds the periodically poled lithium niobate crystal and enhances the SPDC process. The cavity in the idler arm at $1436\,\mathrm{nm}$ filters the redundant modes, while the etalon suppresses the secondary-clusters in the signal arm at $606\,\mathrm{nm}$ (see Fig. 1 of the main text).

We also report the passive losses along the signal and idler optical paths with the purpose of estimating the effective performances of the photon-pair source starting from the detected count-rates.

\begin{table}[htdp]
\begin{center}
\caption{Transmission of the single optical elements, T, and detector efficiencies, $\eta_{d}$, in the signal ($606\,\mathrm{nm}$) and idler ($1436\,\mathrm{nm}$) arm. }
 \label{tab:losses}

\begin{tabular}{|c|c|c|c|}
\hline  Element & T/ $\eta_{d}$ (signal) & $$  & T/ $\eta_{d}$ (idler) \\

\hline

Dichroic Mirror & $> 99\,\%$ & $$&$93\,\%$\\
Glass Plate& $84\,\%$ & $$&$80\,\%$\\
Band Pass Filter& $95\,\%$ &  $$&$$\\
Etalon & $90 \,\%$ & $\eta_{s}$, $\eta_{i}$ &$$\\
Filter Cavity & $$ & $$ &$50\,\%$\\
Fiber & $31\,\%$ & $$&$60\,\%$ \\
Other optical elements & $80\,\%$ & $$& $$ \\
\hline
Cryostat & $75\,\%$ &  $$&$$\\
Fiber & $60\,\%$ &$\eta_{loss}$& $$\\
Memory duty cycle & $50\,\%$ &$$& $$\\
\hline
SPD & $32\,\%$ &$$& $10\,\%$\\

 \hline
 \end{tabular}
\end{center}
\end{table}

\begin{figure}
   \centering
   \includegraphics[width=0.5\textwidth]{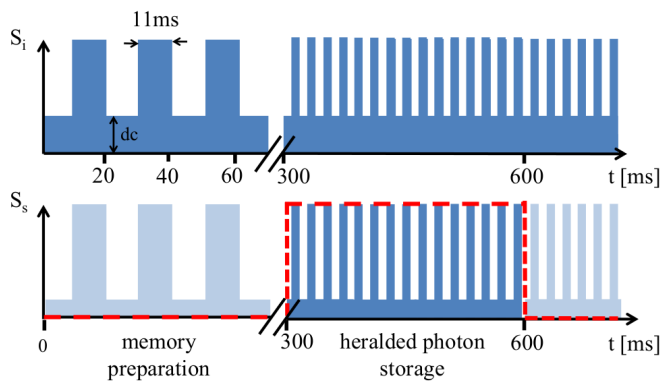}
   \caption{(Color online) Illustration of the source (blue bars) and memory (dotted line) duty cycles.}
   \label{dutycycles}
   \end{figure}

Table \ref{tab:losses} summarizes the transmission of the single optical elements along the path of the signal and the idler  photons at $606\,\mathrm{nm}$ and $1436\,\mathrm{nm}$, respectively.
The total transmission of the signal (idler) photons from the SPDC cavity to the cryostat (SPD) is $\eta_{s} = 18\,\%$ ($\eta_{i} = 22\,\%$).
Figure \ref{dutycycles} shows an illustration of the source and memory duty cycles.
The former, $45\,\%$, derives from the alternation of the SPDC cavity lock and the photon-pair production by means of mechanical choppers. We note that, during the locking period, the measured count rate is the detector dark count. The memory duty cycle, $50\,\%$, is associated with the closure of the mechanical shutter placed after the {\PrYSO} crystal and aimed at protecting the $606\,\mathrm{nm}$ SPD from the eventual leakage of the strong preparation beam into the echo mode. During the preparation time ($300\,\mathrm{ms}$ \cite{Gundogan2013}) we also gate the SPD. While the source duty cycle is the same for the signal and idler photons, the memory duty cycle only affects the signal photons, thus altering the detected coincidence rate, $C_{d}$. For this reason, the transmission of the $606\,\mathrm{nm}$ photons from the cryostat to the signal detector, $\eta_{loss} = 22.5\,\%$, also takes into account the memory duty cycle.

\begin{figure}
   \centering
   \includegraphics[width=.45\textwidth]{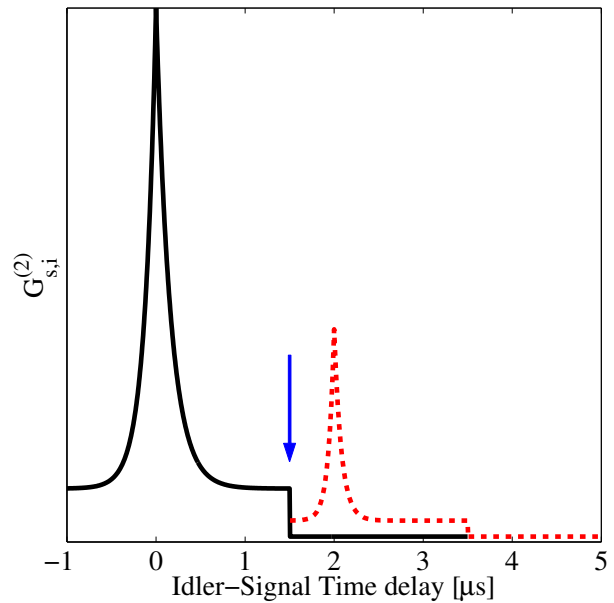}
   \caption{(Color online) Illustration of the $G^{(2)}_{s,i} (t)$ cross-correlation function for the signal photons when passing through the transparency window (black solid curve) and being absorbed and reemitted by the AFC (red dashed curve). The arrow indicates the noise drop due to the pump beam switching off.}
   \label{illustration}
   \end{figure}

During the measurement time, we optimize the detection of the stored and retrieved photons by switching off the pump beam after the detection of an heralding photon. This interrupts the SPDC process and prevents the production of noise due to uncorrelated photons in the region of the coincidence histogram where the AFC echo is expected to occur. Figure \ref{illustration} shows an illustration of the coincidence histograms for the input (solid curve) and the retrieved (dashed curve) photons, where the noise drop is emphasized by an arrow. The time delay between the idler photon detection and the pump turning off changes with the storage time, being modulated in order the noise drop to occur $500\,\mathrm{ns}$ before the absorbed photons are re-emitted. This allows us to measure the noise affecting the echo in an appropriately large window. The pump beam remains off during $20\,\mathrm{\mu s}$. 

\section{Memory characterization}

Before the single photon storage experiments, we characterize our
memory initially with bright pulses and then with weak coherent
states obtained from a laser beam strongly attenuated with a set
of neutral density filters. In both cases we use gaussian pulses
with a duration comparable with the correlation time of the 
photon-pairs produced by the SPDC source, i.e. $108\,\mathrm{ns}$. The
mean number of photons $\mu$ is evaluated from the raw number of
counts detected by the SPD when the AFC is not prepared,
back-propagated taking into account the detector efficiency, the
coupling efficiency into the fiber, and the losses induced by the
crystal when a transparency window is prepared (see Table \ref{tab:losses}).

\begin{figure}
   \centering
   \includegraphics[width=0.45\textwidth]{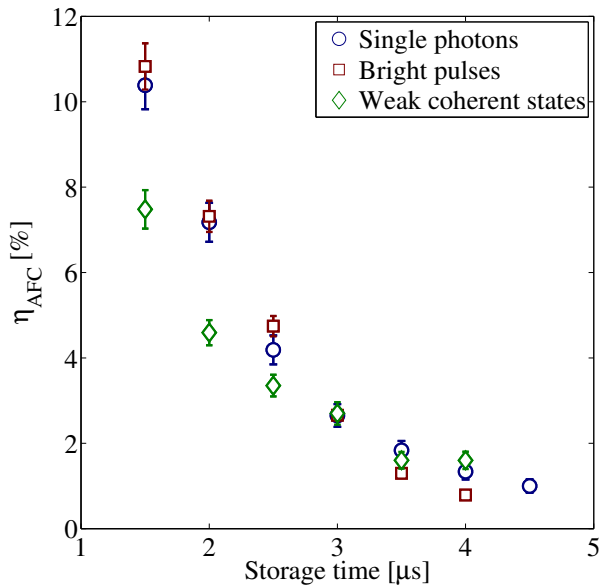}
   \caption{(Color online) AFC echo efficiency as a function of the storage time for bright pulses (squares), weak coherent states (diamonds) and single photons (circles).}
   \label{AFCvsTauS}
   \end{figure}

Fig. \ref{AFCvsTauS} reports the AFC echo efficiency obtained at
increasing storage times for bright pulses (squares), weak
coherent states (diamonds), and single photons (circles). The mean
number of photons in the weak coherent pulses is fixed to $\mu =
0.01$, to be of the same order of magnitude as the heralding efficiency
of the SPDC source. We stress that observing a constant
AFC echo efficiency when moving from bright pulses, obtained from
a stabilized laser beam, to single photons, implies the frequency
jitter of the photon-pair source to be limited and the double
locking system efficient.

In the framework of the theoretical model presented in ref. \cite{Afzelius2009}, the AFC echo efficiency can be expressed as
\begin{equation}
\eta_{AFC} \approx {\tilde{d}}^2 e^{-7/F^2} e^{-\tilde{d}} e^{-d_{0}}, \\
\label{eta_AFC}
\end{equation}

where $F = \Delta / \gamma$ is the finesse of the comb with
$\gamma$ being the peak width, $\tilde{d} = d/F $ is the effective
optical depth experienced by the absorbed photons with $d$ being
the optical depth of the peaks, and $d_{0}$ is the absorbing
background. From the example of AFC shown in Fig. 1(b) of the main
text, we extract the average peak width and separation, optical
density and absorbing background ($\gamma = (76 \pm
9)\,\mathrm{kHz}$, $\Delta = (488 \pm 7)\,\mathrm{kHz}$, $d = 4.9
\pm 0.4$, and $d_{0} = 0.56 \pm 0.03$, respectively) and we
predict $\eta_{AFC} = (13 \pm 3) \,\%$ for a storage time of
$2\,\mathrm{\mu s}$, to be compared with the value of $\eta_{AFC}
= (7.2 \pm 0.5) \,\%$ observed in the heralded single photon
storage (see Fig. \ref{AFCvsTauS}). The disagreement between the expected
and the measured efficiency might be due to a not complete absorption
of the photons due to the difference between
the AFC total width and the photon bandwidth.  It is worth noting, however,
that the comb optical density might be underestimated due to the
limited dynamic range of the detector. This could lead to the overestimation of the
peak width and, consequently, of the theoretical AFC efficiency due to
a suboptimal finesse.
Note also that higher efficiencies could be reached by tailoring
square AFC peaks \cite{Bonarota2010}.

\section{Second-order auto-correlation measurements}

For a pair of independent classical fields, the Cauchy-Schwarz
inequality must be fulfilled \cite{Kuzmich2003}:
\begin{equation}
 R = {\frac{(g^{(2)}_{s,i})^2} {{g^{(2)}_{i,i} \times
g^{(2)}_{s,s}} }} \leq 1
\end{equation}
where $g^{(2)}_{s,s}$ and $g^{(2)}_{i,i}$ are normalized second-order 
auto-correlation functions for signal and idler fields,
respectively. The light emitted by spontaneous down conversion is
usually well approximated by two-mode squeezed states, which
exhibit thermal statistics for the signal and idler fields,
i.e. $g^{(2)}_{s,s}, g^{(2)}_{i,i} = 2$ \cite{Tapster1998}. Under the assumption of
two-mode squeezed states, measuring  $ g^{(2)}_{s,i}>2$ is
therefore sufficient to violate the Cauchy-Schwarz inequality and
give strong evidence of non-classical correlations. However, in
order to obtain a violation of Cauchy-Schwarz inequality without
assumptions on the created state, it is required to measure
directly $g^{(2)}_{s,s}$ and  $g^{(2)}_{i,i}$. We thus assembled a
Hanbury Brown-Twiss set-up, inserting fiber beam-splitters after
the cryostat in the $606\,\mathrm{nm}$ arm (Thorlabs), when only a
transparency window is created, and after the filter cavity in the
$1436\,\mathrm{nm}$ arm (AFW Technologies). Additional SPDs are
employed after the fiber beam-splitters for the detection of the
$606\,\mathrm{nm}$ and the $1436\,\mathrm{nm}$ photons (Count from
Laser Components and id210 from ID Quantique, respectively).

\begin{figure}
   \centering
   \includegraphics[width=.45\textwidth]{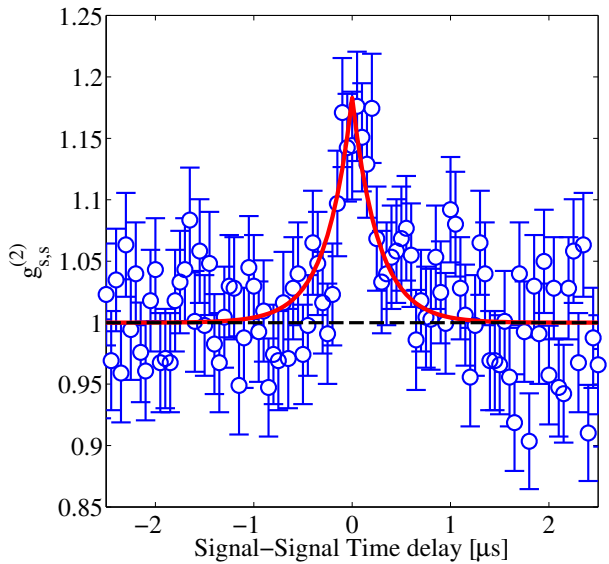}
   \caption{(Color online) Normalized second-order auto-correlation function of the signal photons.
The pump power is $5\,\mathrm{mW}$ and the integration time is
$18\,\mathrm{hours}$. The time-bin size is $ 50\,\mathrm{ns}$. The error bars are
calculated according to Poissonian statistics.}
   \label{g2auto}
   \end{figure}

Figure \ref{g2auto} shows the intensity auto-correlation function
of the signal photons at $606\,\mathrm{nm}$ measured with a pump
power of $5\,\mathrm{mW}$: it exhibits the clear fingerprint of
bunching behavior in the fact that the peak maximum at zero delay
exceeds unity by several standard deviations. The  normalized
second-order auto-correlation measured in a time window $\Delta
t_{d} = 400\,\mathrm{ns}$ is $g^{(2)}_{s,s}(\Delta t_d) = 1.14 \pm
0.02$. Figure \ref{g2auto} also reports the fit of the curve
according to a symmetric exponential decay about zero delay, from
which we extrapolate the decay time $t_{c} = 265\,\mathrm{ns}$ and
the FWHM correlation time $\tau_{ac} = 160\,\mathrm{ns}$.
Analogously, we find for the idler photons a value of
$g^{(2)}_{i,i}(\Delta t_d) = 1.07 \pm 0.01$. The non-classical
nature of the photon correlation is therefore confirmed, at pump
power of $5\,\mathrm{mW}$, by the violation of the Cauchy-Schwarz
inequality in the form $ R =61 \pm 2 \not\leq 1$. We also measure
$g^{(2)}_{s,s}(\Delta t_d)$ and $g^{(2)}_{i,i}(\Delta t_d)$ at a
lower pump power, i.e. $2\,\mathrm{mW}$, leading to higher
violation of the Cauchy-Schwarz inequality $R=142 \pm 10$ (see
Table \ref{tab:g2auto}).

The measured values of $g^{(2)}_{s,s}$ and $g^{(2)}_{i,i}$ are
significantly lower than the value of 2 expected for the thermal
states produced by SPDC phenomena \cite{Tapster1998}. We describe
here some additional measurements which, joined to a simple model
for the noise characterization, provide an
explanation of the disagreement between the experimental and the
expected values.

Exploiting the same Hanbury Brown-Twiss set-up, we perform
auto-correlation measurements of the signal photons when
travelling through the {\PrYSO} crystal without any spectral
feature tailored. The inhomogneously broadened Pr$^{3+}$ profile
should act as a filter for the single photons, thus returning a
flat coincidence histogram between the two arms of the fiber
beam-splitter. However, it gives the measure of any kind of
non-resonant, allegedly broadband, noise affecting the system
\cite{Wolfgramm2011}.

\begin{figure}
   \centering
   \includegraphics[width=.45\textwidth]{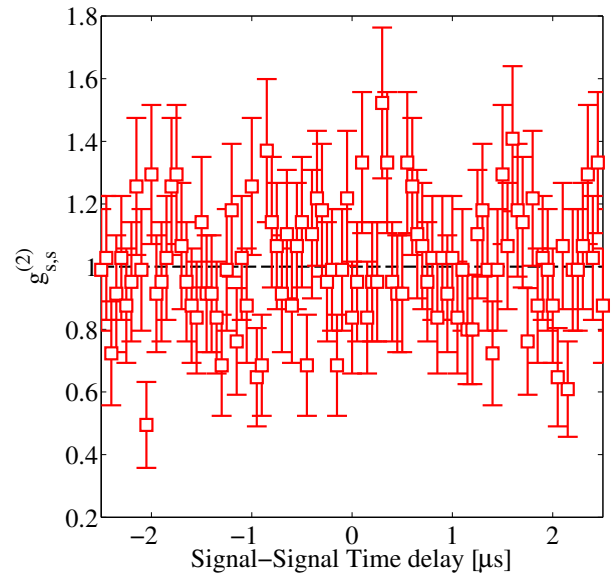}
   \caption{(Color online) Second-order auto-correlation function of signal
photons transmitted by the inhomogeneously broadened line of
Pr$^{3+}$ at $606\,\mathrm{nm}$. The pump power is
$5\,\mathrm{mW}$ and the integration time is $2\,\mathrm{hours}$.
The time-bin size is $ 50\,\mathrm{ns}$. The error bars are
calculated according to Poissonian statistics. }
   \label{g2abs}
   \end{figure}

The red squares in Fig. \ref{g2abs} display the measured
second-order auto-correlation function of the photons transmitted
by the crystal. We note that no evidence of photon bunching is
present, thus ruling out any contribution from the
secondary-clusters. In fact, they are separated from the
main-cluster by a value ($44.5\,\mathrm{GHz}$ \cite{Fekete2013})
significantly higher than the inhomogeneous broadening of the
Pr$^{3+}$ optical transition at $606\,\mathrm{nm}$
($5\,\mathrm{GHz}$ \cite{Gundogan2013}). If they were not
completely suppressed by the etalon in the signal arm, they would
not be absorbed by the crystal and would reach the detector giving
a bunching peak.

A quantitative comparison between the two measurements reported in
Fig. \ref{g2auto} and Fig. \ref{g2abs} can provide hints about the
noise contribution and thus help us to interpret the
experimentally obtained $g^{(2)}_{s,s}(\Delta t_d)$ value.

We define $N_{A} = S_{A} + B_{A}$ as the count-rate at the
detector A when the transparency window is created. $S_{A}$ is the
SPDC contribution and $B_{A}$ the background, inclusive of
uncorrelated noise and detector dark counts. With the assumption
that it is mainly non-resonant with the Pr$^{3+}$ absorption, we
extract the background from the count-rate at the same detector in
the measurement where no transparency pit is tailored.
Analogously, $N_{B} = S_{B} + B_{B}$ is the count-rate at the
detector B.

At a time delay much greater than the correlation time, the number of coincidence counts is proportional to the product
\begin{equation}
N_{A} N_{B} = S_{A}  S_{B}+  S_{A} B_{B} + S_{B} B_{A} + B_{A}  B_{B},\\
\end{equation}

while at zero delay it is proportional to $N_{A} N_{B} + S_{A}
S_{B}$ due to the bunching behaviour of the SPDC photons (assuming
thermal statistics). The proportion in both cases is given by the
product of the window size and the integration time. The expected
value of the intensity auto-correlation function is thus
\begin{equation}
{g_{s,s}^{(2)}}^{\mathrm{th}} (0)= \frac{N_{A} N_{B} + S_{A} S_{B}}{N_{A} N_{B}} = 1 + \frac{S_{A}  S_{B}}{N_{A} N_{B}}.\\
\label{g2autoteo}
\end{equation}

For $P_p= 5\,\mathrm{mW}$ pump power, we measure background to
signal ratios $B_{A}/S_{A} = 0.85$ and $B_{B}/S_{B} = 0.98$. This
leads to  value of ${g_{s,s}^{(2)}}^{\mathrm{th}} (0) = 1.27 $.  This
result suggests that the noise affecting our system is
prevailingly non-resonant with the Pr$^{3+}$ absorption.

The inferred ${g_{s,s}^{(2)}}^{\mathrm{th}}(0)$ refers to the ideal case of
an infinitely small integration window about zero delay, thus it
is not directly comparable to the experimental value
$g^{(2)}_{s,s}(\Delta t_{d}) = 1.14 \pm 0.02$, which is measured
in a finite time window $\Delta t_{d} = 400\,\mathrm{ns}$.
To take this effect into account, we consider a
symmetric exponential decay with decay time $t_c$ for the
correlation function and find
\begin{equation}
{g^{(2)}_{s,s}}^{\mathrm{th}} (\Delta t_d)={g_{s,s}^{(2)}}^{\mathrm{th}}(0)
\frac{2t_c}{\Delta t_d}(1-e^{(-\frac{\Delta t_d}{2t_c})})+1,
\end{equation}

where the factor of two takes into account the symmetric
exponential. For the fitted $t_c=265 \,\mathrm{ns}$, we find
${g_{s,s}^{(2)}}^{\mathrm{th}}(\Delta t_d) = 1.19$.

We also perform a similar analysis for a pump power of 2 mW. Table
\ref{tab:g2auto} summarizes the experimental and theoretical
values for the different powers of the $426.2\,\mathrm{nm}$ pump.

\begin{table}[htdp]
\begin{center}
\caption{Normalized intensity auto-correlation values of signal
and idler photons as obtained from the experimental histogram and
from the theoretical model for different powers of the
$426.2\,\mathrm{nm}$ pump. The R values, as calculated from the
experimental values of $g^{(2)}_{s,s}$, $g^{(2)}_{i,i}$, and
$g^{(2)}_{s,i}$, are also reported. All values are for a detection
window $\Delta t_d= 400\, \mathrm{ns}$.}
\label{tab:g2auto}

\begin{tabular}{|c|c|c|}
\hline $P_p $ & $2\,\mathrm{mW}$ &  $5\,\mathrm{mW}$ \\

\hline

$g_{s,s}^{(2)}$ & $1.09 \pm0.04$ & $1.14 \pm 0.02$ \\
${g_{s,s}^{(2)}}^{\mathrm{th}}$&$1.15$&$1.19$ \\
$g_{i,i}^{(2)}$&$1.03 \pm 0.01$&$1.07 \pm 0.01$ \\
${g_{i,i}^{(2)}}^{\mathrm{th}}$&$1.003 $&$1.03  $ \\
$g_{s,i}^{(2)}$&$12.6 \pm 0.5$&$8.7 \pm 0.2$ \\
$R$&$142 \pm 10$&$61 \pm 2$ \\

 \hline
 \end{tabular}

\end{center}

\end{table}

The interpretation of the ${g_{i,i}^{(2)}}$ value measured for the
idler photons deserves a separated discussion. In fact, in this
case the background contribution cannot be directly measured as
for the signal photons. We note, however, that, due to the
presence of the filter cavity in the idler mode and to much higher
dark count rates for the SPDs, the background for the infrared
photons is dominated by the dark counts of the detectors (400 Hz
and 1200 Hz for the id220 and id210, respectively). Applying Eq.
\ref{g2autoteo} to the idler photons, where $B_{A}$ and $B_{B}$
are evaluated from a dark count measurement, we calculate an
expected value of ${g_{i,i}^{(2)}}^{\mathrm{th}}(\Delta t_d) \approx 1.03$
for $5\,\mathrm{mW}$ pump power, in reasonable agreement with the
experimental ${g_{i,i}^{(2)}(\Delta t_d)} = 1.07 \pm 0.01$.

\section{Linear dichroism measurements}

\begin{figure}
   \centering
   \includegraphics[width=.45\textwidth]{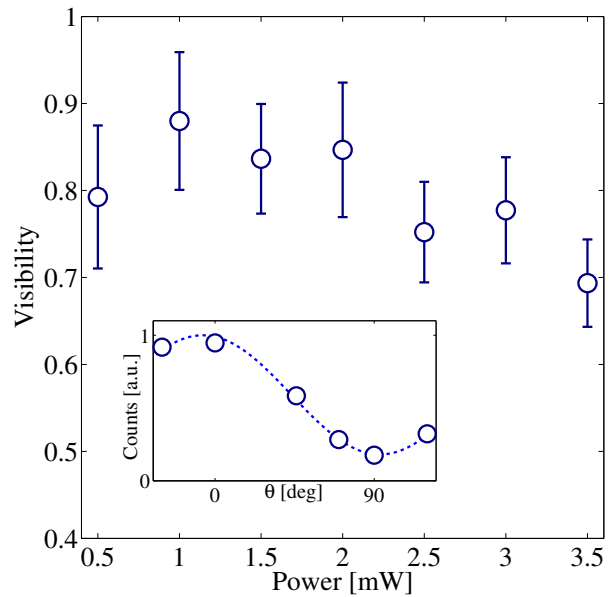}
   \caption{(Color online) Visibility as a function of the pump power as obtained from coincidence histograms measured with different polarization of the input photons when no transparency window is created. The error bars are calculated according to Poissonian statistics. Inset: Normalized coincidence counts in a $400\, \mathrm{ns}$ window as a function of the polarization of the $606\,\mathrm{nm}$ photons travelling through the crystal. The pump power is $3.5\,\mathrm{mW}$. }
   \label{VvsP}
   \end{figure}

To investigate the disagreement between the values for the input and the stored photons of the normalized second-order cross-correlation between signal and idler (see Fig. 2 of the main text), $g^{(2)}_{s,i}$, we perform linear dichroism measurements with the heralded photons transmitted by the inhomogeneous line of the {\PrYSO} crystal. We collect the $G^{(2)}_{s,i} (t)$ cross-correlation function between signal and idler for different polarizations of the signal photons without preparing either the AFC or the transparency window. From the coincidence histograms we calculate the $g^{(2)}_{s,i}$ values as a function of polarization, considering a time window of $400\, \mathrm{ns}$ around the peak at zero delay. An example is given in the inset of Fig. \ref{VvsP} in the case of $3.5\,\mathrm{mW}$ pump power. 

The main frame of Fig. \ref{VvsP} displays the visibility as a function of the pump power, extracted from the polarization dependence of the coincidences.
For pump powers up to $2\,\mathrm{mW}$ the conditioned visibility approaches $90\,\%$ and it decreases below $70\,\%$ when the pump power reaches $3.5\,\mathrm{mW}$. This is a further proof of the existence of a non-resonant noise, whose relative weight with respect of the resonant signal is increasing at higher pump powers. From the measured visibilities we can extract the probability to detect coincidences due to photons not resonant with the Pr$^{3+}$ inhomogeneous absorption line, $P_{nr}$, with respect of that of detecting coincidences due to resonant ones, $P_{r}$.

With $\theta$ the angle between the light polarization and the
optical D$_1$ axis of the {\YSO} crystal, we assume that the
probability to detect a coincidence behaves as
\begin{equation}
P(\theta) = P_{nr} + P_{r} \times e^{-OD(\theta)},
\label{visibility}
\end{equation}
where $OD(\theta)$ is the optical density of the Pr$^{3+}$
absorption at that polarization. Taking into accout OD=1.4
measured in the weakly absorbing polarization (D$_1$) and OD=6.9
for the polarization parallel to the D$_2$ axis, we evaluate the
non-resonant coincidences to be about $1 - 2\,\%$ of the resonant
ones for visibilities varying between $90$ and $80\,\%$, which is
the case for pump powers up to $2\,\mathrm{mW}$. As soon as the
visibility drops to values below $70\,\%$, the non-resonant to
resonant photons ratio exceeds $5\,\%$.

\begin{figure}
   \centering
   \includegraphics[width=.45\textwidth]{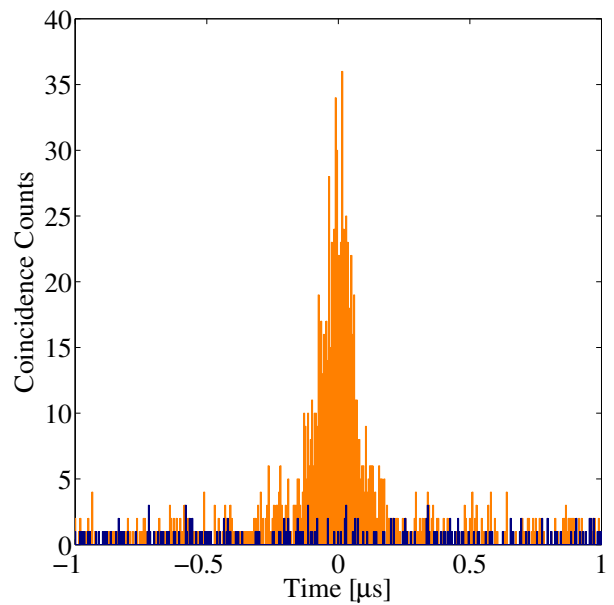}
   \caption{(Color online) $G^{(2)}_{s,i} (t)$ functions acquired at  $2\,\mathrm{mW}$ pump power with the signal photons transmitted by the Pr$^{3+}$ inhomogeneous absorption line (blue trace) and travelling through a transparency window (orange trace).}
   \label{histogram}
   \end{figure}

The described approach enables quantifying the ratio of signal
photons resonant with the Pr$^{3+}$ absorption line with respect 
of the total photons produced, i.e. $97\,\%$ for pump powers up to $3\,\mathrm{mW}$ and
$95\,\%$ for higher pump powers. We note,
however, that a comparison between the coincidence rates measured
in a time window $\Delta t_{d} = 400\,\mathrm{ns}$ for the
$606\,\mathrm{nm}$ photons passing through the absorption profile
and the transparency window (examples at $P_{p} = 2\,\mathrm{mW}$
are shown in Fig. \ref{histogram}), allows us to conclude that the
majority of the photons resonant with the Pr$^{3+}$ inhomogeneous
line are also resonant with the transparency window where the AFC
is tailored. 

As a matter of fact, we find for the photons
transmitted by the inhomogeneous line of  Pr$^{3+}$ a coincidence
rate of $C_{d} = (0.04\pm0.1)\,\mathrm{Hz/mW}$, i.e. less than $5\, \%$ of
the detected coincidence rate of the input photons ($C_{d} =
(0.83\pm0.14)\,\mathrm{Hz/mW}$). In other words, more than $95\,\%$ of the heralded
photons detected after the crystal are resonant with the transparency window.
The non-resonant noise is expected to affect the $g^{(2)}_{s,i}$ values of the input but not those of the stored photons. In fact, the AFC comb selects only a portion of the input photons and returns it in a separated temporal mode, which does not include any uncorrelated noise due to the pump beam switching off after the detection of an idler photon (see Fig. \ref{illustration}).

We now describe a simple picture explaining how, even a small contribution of non-resonant noise, can significantly affect the $g^{(2)}_{s,i}$ values. We can write the normalized intensity cross-correlation in terms of resonant and non-resonant contributions to the coincidence probability as
\begin{equation}
{{g}_{s,i}^{(2)}}^\mathrm{{th}} = \frac{p_{s,i}}{p_{s}p_{i}} = \frac{P_{r} +P_{nr}}{P_{n} +P_{nr}} ,\\
\label{g2res_nonres}
\end{equation}
where $P_{n}$ is the probability of having accidental coincidences due to resonant photons.
Defining $ r = P_{nr}/P_{r}$, Eq. \ref{g2res_nonres} becomes

\begin{equation}
{{g}_{s,i}^{(2)}}^{\mathrm{th}} = \frac{P_{r}}{P_{n}} \times \frac{1 + r}{1 + r(P_{r}/P_{n})} .\\
\label{g2input_echo}
\end{equation}

\begin{table}[htdp]
\begin{center}
\caption{Normalized intensity cross-correlation between signal and
idler photons as obtained from the measurements with the signal
photons passing through the transparency window, $g_{s,i}^{(2)}
(\mathrm{input})$, compared to that estimated,
${{g}_{s,i}^{(2)}}^{\mathrm{th}}$, starting from the cross-correlation of
the retrieved photons, $g_{s,i}^{(2)} (\mathrm{echo})$, for
different powers of the $426.2\,\mathrm{nm}$ pump. }
\label{tab:g2predict}

\begin{tabular}{|c|c|c|}
\hline $P_p $ & $2\,\mathrm{mW}$ &  $3\,\mathrm{mW}$ \\

\hline

$g_{s,i}^{(2)} \mathrm{(echo)}$ & $17.4 \pm2.7$ & $15 \pm 2$ \\
${{g}_{s,i}^{(2)}}^{\mathrm{th}}$&$13 \pm 5$&$10 \pm3$ \\
$g_{s,i}^{(2)} \mathrm{(input)}$&$12.6 \pm 0.5$&$11.7 \pm 0.4$ \\
$r$& $(2.1 \pm 0.4) \,\%$ & $(3.4 \pm 0.5) \,\%$\\

 \hline
 \end{tabular}

\end{center}

\end{table}

The ratio $P_{r}/P_{n}$ is the normalized second-order
cross-correlation after the filtering of any non-resonant
contribution. In a picture where the AFC storage returns an echo
free of non-resonant photons, it is exactly the $g^{(2)}_{s,i}$
value of the echo. We thus apply the proposed model to predict the
$g^{(2)}_{s,i}$ value of the input photons from that of the echo.
The measured and predicted values are summarized in Table
\ref{tab:g2predict} for $2\,$ and $3\,\mathrm{mW}$ pump power.
The prediction of the model is in agreement with the measured
values, which confirms that the AFC acts as a temporal filter for
non resonant noise \cite{McAuslan2012}.

\bibliographystyle{prsty}

\end{document}